\documentclass[prl,reprint,twocolumn,amssymb,nobibnotes,aps,floatfix]{revtex4-1}
\usepackage{graphicx}
\usepackage{verbatim}
\usepackage{amsmath}
\usepackage{multirow}
\usepackage{dcolumn}

\begin{document}

\title{Single Crystal Sapphire at milli-Kelvin Temperatures: Observation of Electromagnetically Induced Thermal Bistability in High Q-factor Whispering Gallery Modes}%

\author{Daniel L. Creedon}%
\email[Electronic address: ]{creedon@physics.uwa.edu.au}
\author{Michael E. Tobar}%
\author{Jean-Michel \surname{Le Floch}}%
\affiliation{University of Western Australia, School of Physics (M013), 35 Stirling Hwy, Crawley WA 6009, Australia}
\author{Yarema Reshitnyk}%
\author{Timothy Duty}%
\affiliation{University of Queensland, School of Mathematics \& Physics, St. Lucia QLD 4072, Australia}
\date{August 6, 2010}%

\begin{abstract}
Resonance modes in single crystal sapphire ($\alpha$-Al$_2$O$_3$) exhibit extremely high electrical and mechanical Q-factors ($\approx 10^9$ at 4K), which are important characteristics for electromechanical experiments at the quantum limit. We report the first cooldown of a bulk sapphire sample below superfluid liquid helium temperature (1.6K) to as low as 25mK. The electromagnetic properties were characterised at microwave frequencies, and we report the first observation of electromagnetically induced thermal bistability in whispering gallery modes due to the material $T^3$ dependence on thermal conductivity and the ultra-low dielectric loss tangent. We identify ``magic temperatures'' between 80 to 2100 mK , the lowest ever measured, at which the onset of bistability is suppressed and the frequency-temperature dependence is annulled. These phenomena at low temperatures make sapphire suitable for quantum metrology and ultra-stable clock applications, including the possible realization of the first quantum limited sapphire clock.
\end{abstract}

\maketitle
Experiments to couple superconducting qubits based on Josephson junctions to microwave resonators ($Q\approx 10^4$) at cryogenic temperatures have been well represented in recent scientific literature for a diverse range of circuit quantum electrodynamic applications. This includes generating non-classical states of microwave cavities such as Fock states, where the limit on producing these non-classical fields is due to the finite photon lifetime (or linewidth) of the resonator, as well as detecting a nanomechanical resonator at or near the ground state \cite{RocheleauNature,OconnellNature,HofheinzNature,HofheinzNature2,Osborn07,DutySuperconducting,Castellanos07}.  Sapphire resonators are of particular interest for future experiments due to their extremely low loss, with electronic $Q$-factors of order $10^9$ at 1.8K \cite{LuitenBook,Hartnett2006apl}, and mechanical $Q$-factors as high as $5 \times 10^8$ at 4.2K \cite{lockeparametric,systemssmalldissipation}. The thermal, mechanical, and bulk electronic properties of sapphire have been characterised extensively over a wide range of temperatures from room temperature to superfluid liquid helium (1.6K) using WG mode techniques \cite{JerzyMTT,JerzyMST}, but have never been examined in the regime approaching the absolute zero of temperature.  Sapphire resonators at millikelvin temperature have the potential to play an important role in the next generation of quantum electronics and metrology experiments by virtue of this anomalously high $Q$-factor. A significant body of research already exists in which sapphire has been used at cryogenic temperatures as a parametric transducer in an effort to reach the Standard Quantum Limit \cite{lockeparametric,lockeparametric2,tobarparametric,lockeparametric3,TobarSQL,Cuthb}. Oscillators can be prepared in their quantum ground state due to very low thermal phonon occupation when $T << hf/{k_B}$, where $h$ and $k_B$ are Planck's and Boltzmann's constants respectively. For microwave oscillators such as those based on single-crystal sapphire resonators, the corresponding temperature regime is in the experimentally accessible millikelvin range, making them ideal candidates for quantum measurement experiments. It is thus important to characterise such devices in this unexplored ultra-low temperature regime.
\begin{figure}[bh]
\includegraphics[width=86mm]{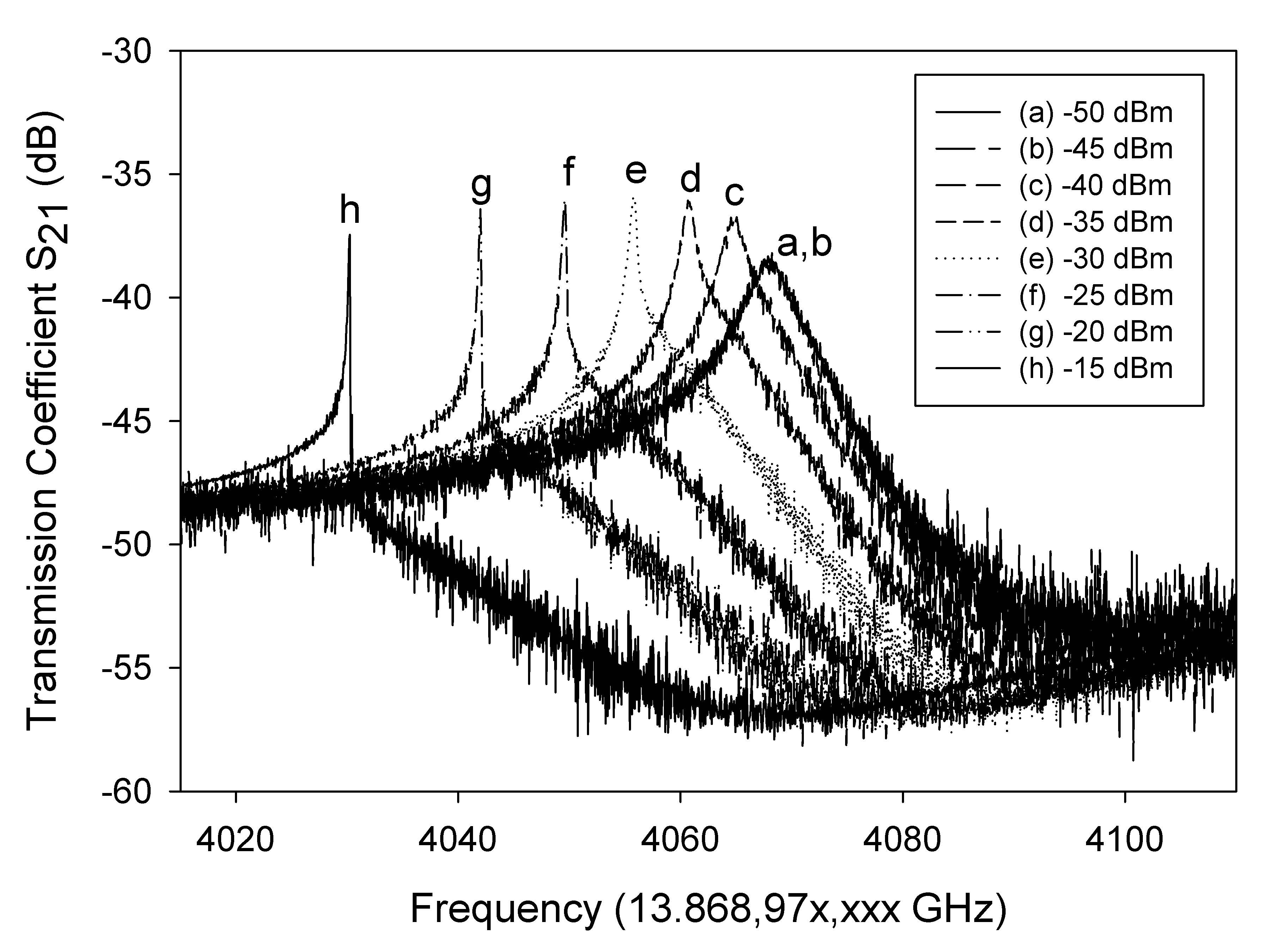}
\caption{\label{fig1}Network analyser measurement of the WGH$_{20,0,0}$ mode in transmission at 50 mK. The excitation power was varied in steps of 5dB from -50dBm to -15dBm, and the observed mode frequency was downshifted. The -50dBm and -45dBm curves are the highest in frequency and lie on top of one another. The threshold power is defined to be the power incident on the resonator which is sufficient to shift the mode frequency by one bandwidth from the `unperturbed' lowest power measurement}
\end{figure}
In this Article we report on the first measurements of the electromagnetic properties of a single-crystal sapphire resonator at millikelvin temperature. The resonator used was a highest purity HEMEX-grade sapphire from Crystal Systems, similar to that used in several Cryogenic Sapphire Oscillator (CSO) \cite{TobarMann,Hartnett2006apl,Locke2006rsi} and Whispering Gallery Maser Oscillator (WHIGMO) experiments \cite{Pyb2005apl,Benmessai2008prl,Creedon2010,Benmessai2007el}.  Here, we report on the observation of the lowest frequency-temperature turning points for Whispering Gallery (WG) mode resonances ever measured, as well as making the first observation of a thermal bistability effect in sapphire for this ultra-low temperature regime. We give a model to predict thermal bistability threshold power and show that the effect is dependent on the thermal conductivity of the sapphire. Furthermore, we show that the bistability effect may be suppressed by operating at a ``magic temperature'', where a frequency-temperature turning point occurs. Thus, combining the low temperature operation with stabilisation at a millikelvin frequency-temperature turning point gives the potential to realise a sapphire based frequency standard at the quantum limit (rather than the thermal limit which has been previously reported \cite{Benmessai2008prl}).\\
\begin{figure}[b]
\includegraphics[width=86mm]{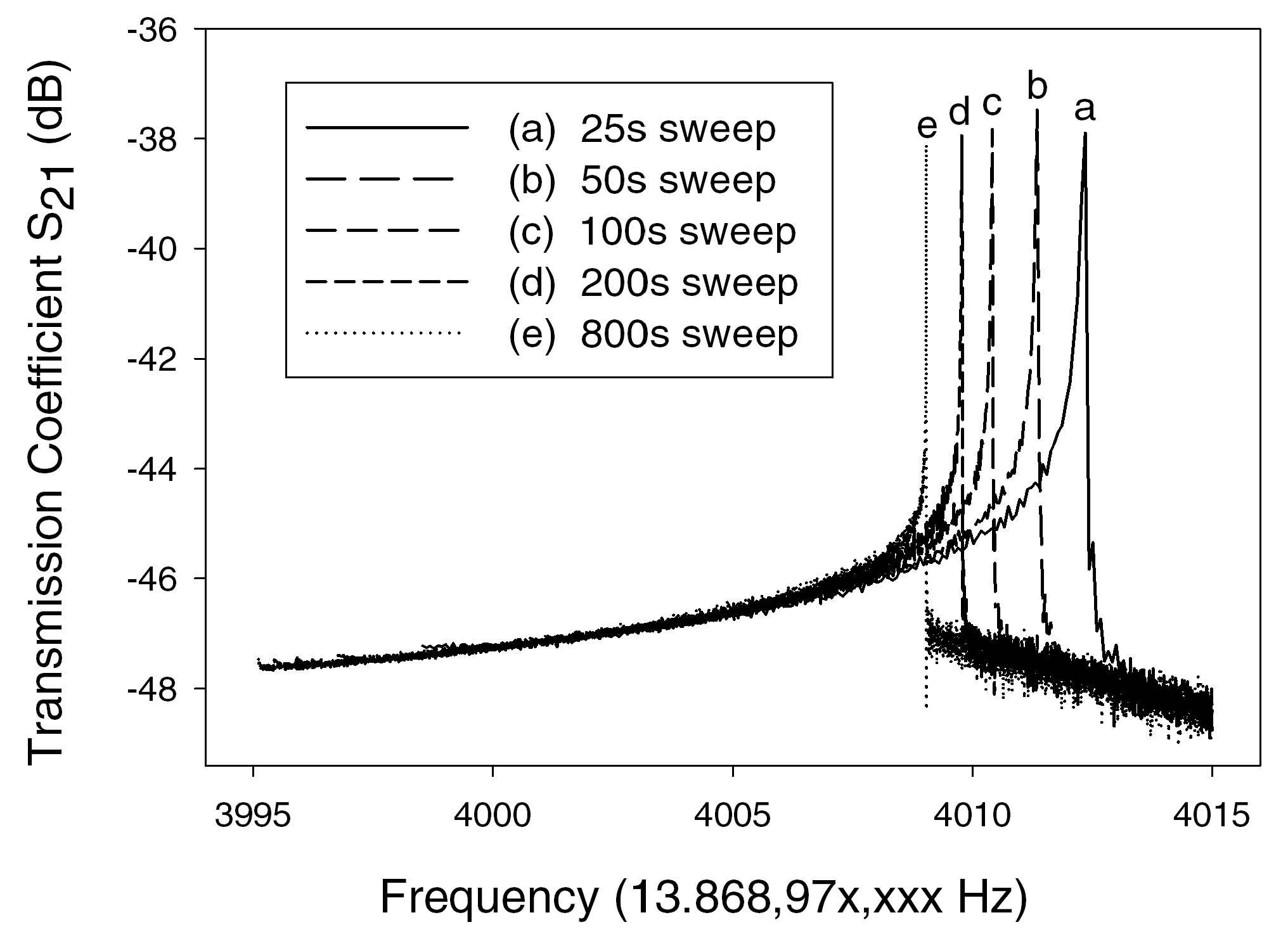}
\caption{\label{sweeptime}The WGH$_{20,0,0}$ mode in transmission for varying sweep time. Sweep direction is increasing in frequency in all cases. The linewidth of the mode narrows sharply with sweep time due to temperature dependence of permittivity of the sapphire. The governing equations for the lineshape are given in \cite{Vahala04}}
\end{figure}
The sapphire resonator, a cylinder 5cm diameter $\times$ 3cm height, was cleaned in acid and mounted in a silver plated copper cavity. The resonator is machined such that the anisotropy c-axis of the sapphire is aligned with the cylindrical z-axis. A radially oriented loop probe and axially oriented straight antenna were used to couple microwave radiation in and out of the crystal. The cavity was attached to the mixing chamber of a dilution refrigerator with a copper mount and cooled to 25mK. The fundamental quasi-transverse magnetic Whispering Gallery modes WGH$_{m,0,0}$, with azimuthal mode number $m$ from 13 to 20, were characterised over a range of temperatures using a vector network analyser. We observed that particularly high-$Q$ WG modes exhibited a hysteretic behaviour, which was thermal in nature. The frequency of the WG modes supported in the resonator are dependent on both the physical dimensions of the crystal and its permittivity, the latter effect being more than an order of magnitude stronger \cite{systemssmalldissipation}. As the network analyser sweeps in frequency, heating occurs as power is deposited into the sapphire on resonance. The change in permittivity due to temperature causes a shift in the resonant frequency of the mode in the opposite direction to the sweep. The result is an astoundingly narrow, yet artificial linewidth with a sharp threshold. If the frequency was swept in the opposite direction, the mode frequency is shifted in the same direction as the sweep, and an artificially broadened linewidth would be observed.
A similar effect in such dielectric resonators has only been observed at optical frequencies in fused silica microspheres \cite{europhys}. It was shown that the ``thermal bistability'' caused either narrowing or broadening of the line resonance depending on the direction of the frequency sweep during measurement. Examples of optical bistability are numerous in the literature \cite{Braginsky,Vahala04,VahalaAPL}, but are normally attributed to a $\chi^{(3)}$ Kerr nonlinearity, which results in a threshold power for optical bistability that scales like $Q^{-2}$. Collot et al. \cite{europhys} note that for mode Q-factors below $10^9$ the thermal bistability effect dominates over the Kerr effect due to significantly lower threshold power. For quality factors in the range of $10^9$, the effects can be distinguished by the observed dependence of threshold power on Q. We find an excellent fit using a thermal model (see Eqn.\ref{eq3} and Fig.\ref{thresh}) which has a threshold power that scales like $Q^{-1}$, showing that the effect is clearly thermal in nature.
A measurement of the WGH$_{20,0,0}$ mode was made (see Fig. \ref{fig1}) which shows the first observation of this thermal bistability effect in a millikelvin sapphire resonator.  A sharp threshold was observed, giving an FWHM linewidth of only 0.00173 Hz, which was strongly dependent on input power. Our experimental apparatus was unable to sweep downwards in frequency, but the bistability effect is still observable by varying the sweep speed. Figure \ref{sweeptime} shows the effect of the thermal bistability for a range of sweep times. Note that only the resonant peak moves; the longer sweep time results in more time spent per measurement point, depositing more power into the resonator and creating a larger apparent frequency shift and linewidth change. As sufficient heat is deposited into the resonator only on resonance, the off-resonance transmission does not depend on sweep time.
\begin{figure}[tb]
\includegraphics[width=86mm]{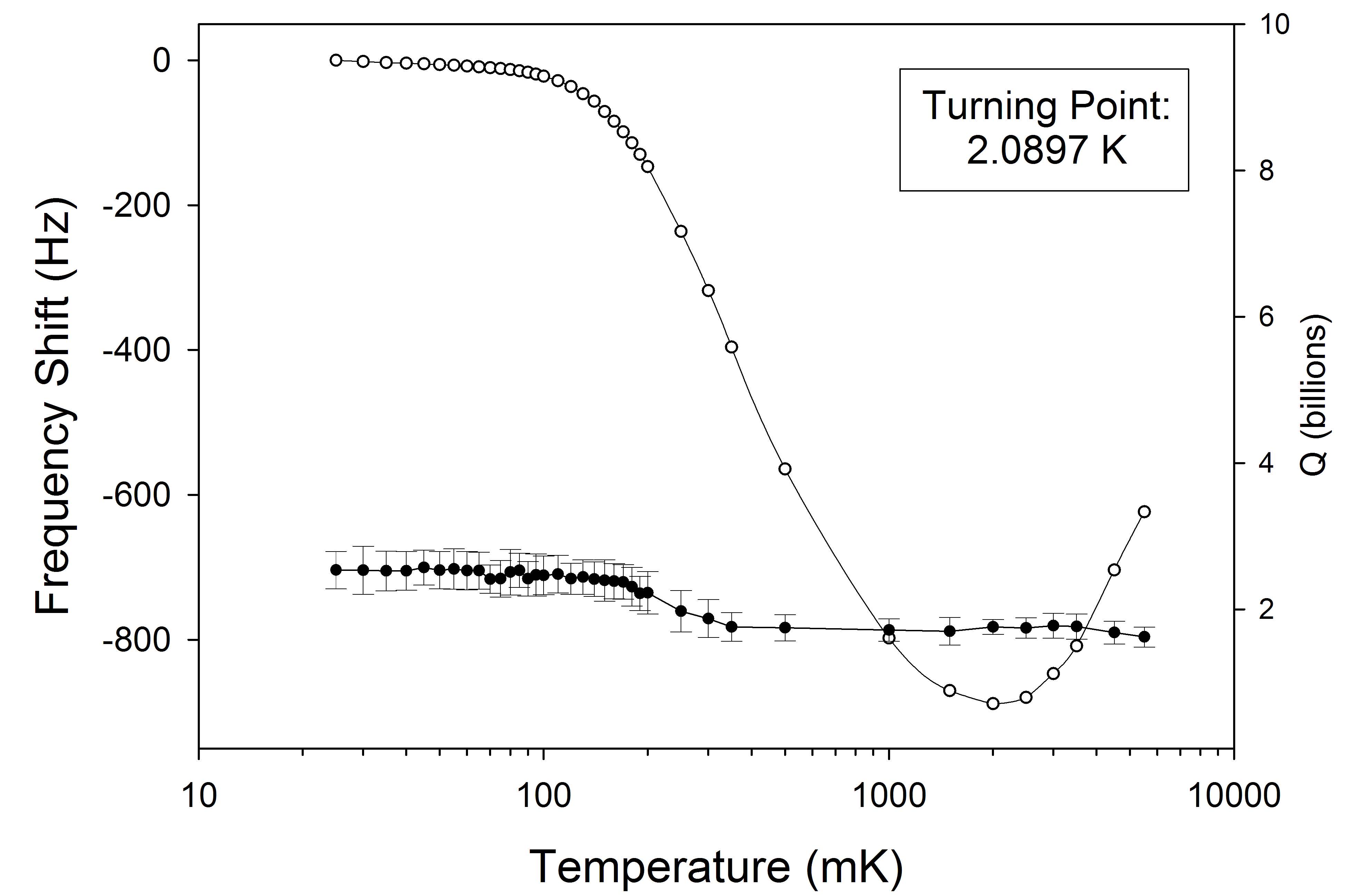}
\caption{\label{fig3}Temperature dependence of $Q$-factor (shaded circles) and frequency (empty circles) for the 20$^{\text{th}}$ azimuthal order WG mode. Note that due to paramagnetic impurities in the sapphire, ${df}/{dT}$ is annulled at 2.0897K. The $Q$-factor remains approximately constant over a wide range of temperature.}
\end{figure}
\begin{table}[b]
\caption{\label{alphaTable}Measured parameters for the thermal coefficient and Q}
\begin{tabular}{|c|c|c|c|}
\hline
$\boldsymbol{m}$	&	$\boldsymbol{T}$ \textbf{(mK)}	&	$\boldsymbol{Q}$		&	$\boldsymbol{\alpha}$ \\
\hline
\hline
\multirow{2}{*}{14}	&	100	&	$7.84 \times 10^8$	&	$-6.66 \times 10^{-9}$ \\ \cline{2-4}
		& 800	&	$1.7 \times 10^9$	&	$-1.3 \times 10^{-11}$  \\
\hline
	 	& 100	&	$6.17 \times 10^8$	&	$-6.85 \times 10^{-8}$ 	\\ \cline{2-4}
	19	& 440	&	$5.40 \times 10^8$	&	$-1.10 \times 10^{-7}$	\\ \cline{2-4}
	 	& 2260	&	$5.85 \times 10^8$	&	$-1.60 \times 10^{-11}$  \\
\hline
		& 200	&	$2.21 \times 10^9$	&	$-1.19 \times 10^{-7}$  	\\ \cline{2-4}
	20	& 630	&	$1.72 \times 10^9$	&	$-3.26 \times 10^{-8}$  	\\ \cline{2-4}
		& 2100	&	$1.76 \times 10^9$	&	$-5.85 \times 10^{-11}$	\\
\hline
\end{tabular}
\end{table}
It is possible to model the threshold power at which bistable behaviour becomes apparent. Considering a temperature dependent fractional frequency shift of the WG mode of interest, $\Delta{\nu}/\nu = -\alpha \Delta{T}$, where the temperature coefficient $\alpha$ is experimentally determined, we then expect thermal bistability for a threshold temperature rise of the resonator $\Delta{T_{th}}= \frac{1}{\alpha} \frac{\Delta{\nu}}{\nu} = \frac{1}{\alpha Q}$ K. An expression for the threshold power required to achieve this bistability is given by \cite{europhys}:
\begin{equation}
\label{powerthreshold}
P_{th} = \frac{C_{p} \rho V_{eff} \Delta T_{th}}{\tau_{\text{T}}}
\end{equation}
where $C_{p}$ is the heat capacity of sapphire, $\rho$ is its density (4.0 g/cm$^3$ \cite{handbookchemphys}), $\tau_{T}$ the characteristic heat diffusion time constant, and $V_{eff}$ the effective volume occupied by the Whispering Gallery mode. The heat diffusion time constant in turn may be expressed as:
\begin{equation}
\label{timeconstant}
\tau_{\text{T}} = \frac{l m C_{p}}{A k}
\end{equation}
where $l$ and $A$ are the length and cross-sectional area of the sapphire segment, $m$ is the mass of the sapphire, and $k$ the thermal conductivity \cite{tobarifcs1994}. Finally, an expression (independent of the heat capacity and thermal time constant) is derived by combining Equations \ref{powerthreshold} and \ref{timeconstant}:
\begin{equation}
\label{eq3}
P_{th} = \frac{A k}{l \alpha Q} \frac{m_{eff}}{m}
\end{equation}
Where $m_{eff}$ is the mass of the effective volume occupied by the WG mode. The expression for the threshold power is now clearly only a function of the thermal coefficient $\alpha$, $Q-$factor, and thermal conductivity of the sapphire, as well as its dimensions. The thermal conductivity $k$ was estimated by fitting an approximate $T^3$ power law to extrapolate below 1K from the data for sapphire in Touloukian et al.\cite{touloukian}, giving $k=0.039T^{2.8924}$ W cm$^{-1}$K$^{-1}$. The thermal time constant of sapphire remains similar to that at liquid helium temperature because the heat capacity follows a similar cubic law to the thermal conductivity, leaving the ratio unchanged (Eqn. \ref{timeconstant}). However, the threshold for bistability is substantially lowered with respect to liquid helium temperature due to the reduction in thermal conductivity of the sapphire.\\
To experimentally determine the thermal coefficient $\alpha$, frequency measurements of several Whispering Gallery modes were made over a range of temperatures from 25mK to 5.5K. The modes examined were fundamental quasi-transverse magnetic modes WGH$_{14,0,0}$, WGH$_{19,0,0}$ and WGH$_{20,0,0}$. The modes were excited using a vector network analyser at low power, typically $-45$ dBm. In this way, saturation of residual paramagnetic spins in the sapphire was avoided.  The temperature of the resonator was controlled at a number of points between 25-5500mK using a Lakeshore Model 370 AC Resistance Bridge, and custom data acquisition software recorded the temperature, $Q$-factor, and frequency of the modes in transmission. The base temperature of the dilution refrigerator was 23mK, and temperature control was stable to within several millikelvin. The temperature dependence of mode frequency was mapped to produce plots such as Fig. \ref{fig3}, and several temperatures were chosen to measure the threshold power at which thermal bistability became apparent. The temperatures were chosen to reflect a range of values for the thermal coefficient $\alpha$, ranging from nominally zero near the frequency-temperature turnover point, to a maximum at the largest slope. As previously, we define the threshold power to be the power incident on the resonator which is sufficient to shift the mode frequency by one bandwidth from the unperturbed low power measurement. Equation \ref{eq3} is then used to calculate the theoretical threshold power. The experimentally determined thermal coefficients are summarised in Table \ref{alphaTable}, and a particular example of the measured and calculated threshold power is given in Fig. \ref{thresh} for WGH$_{20,0,0}$.\\
\begin{figure}[tb]
\includegraphics[width=86mm]{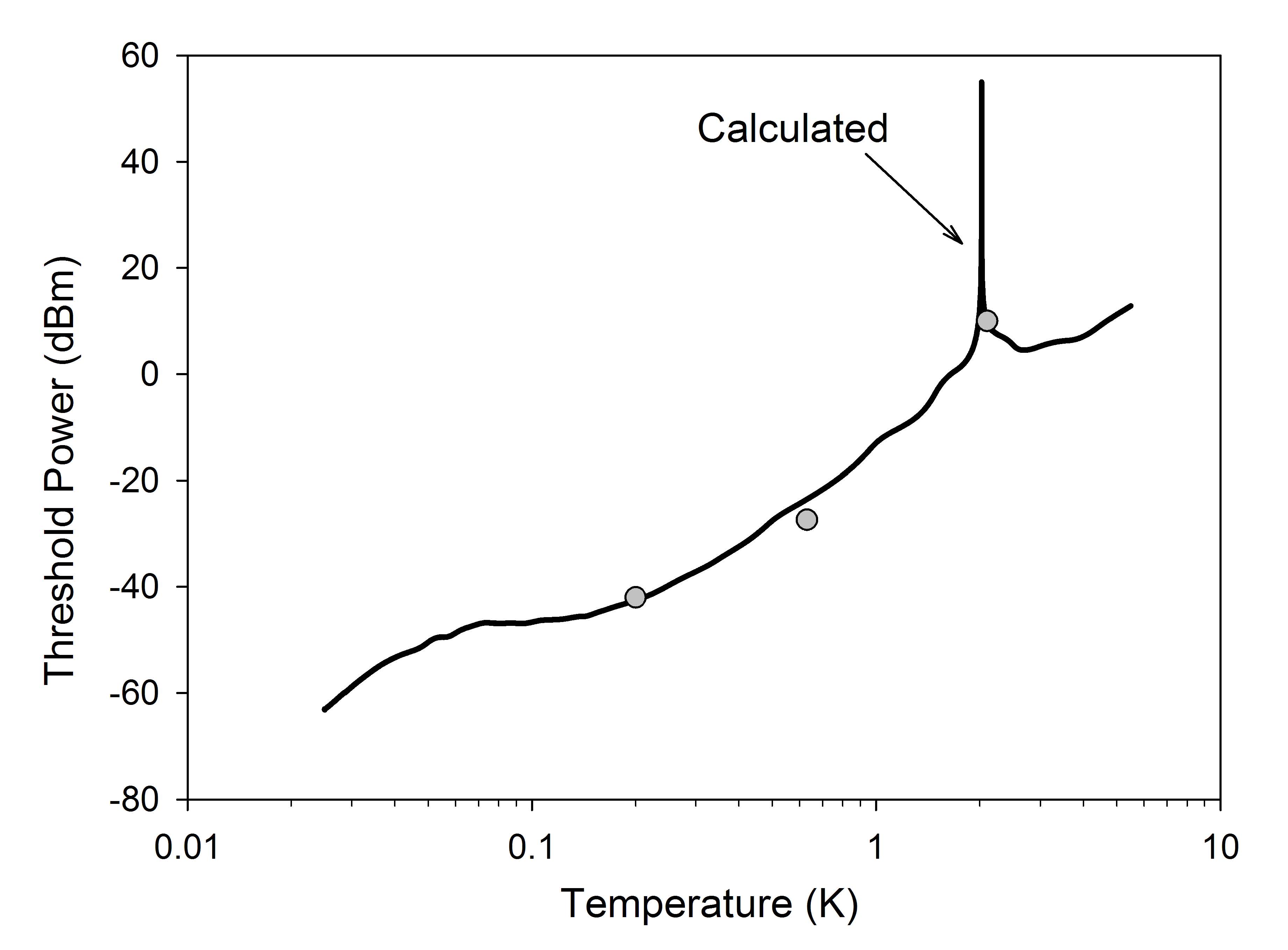}
\caption{\label{thresh}Predicted threshold power (using Eqn.\ref{eq3}) as a function of temperature for WGH$_{20,0,0}$. Shaded circles are the measured threshold values for 200, 630, and 2100 mK.}
\end{figure}
Clearly, operation at the frequency temperature turning point is advantageous as the thermal coefficient $\alpha$ approaches zero and the threshold power required for bistability approaches infinity. Temperature turning points have been observed with no bistability at input powers up to 20 dBm from 4-9K in similar sapphire resonators, and are caused by residual paramagnetic impurities such as Ti$^{3+}$, Cr$^{3+}$, Mo$^{3+}$, V$^{3+}$, Mn$^{3+}$, and Ni$^{3+}$ present at concentrations of parts-per-billion to parts-per-million. The opposite sign effects of temperature-dependent Curie law paramagnetic susceptibility, and temperature dependence of permittivity\cite{DickWang1,JonesBlair1988el,Mann1992jpDap,HartnettTi3,TobarJPhysD} cause a turnover in the frequency-temperature dependence. Operating at this turning point or ``magic temperature'' allows frequency fluctuations due to temperature instability to be annulled to first order, and has been crucial to achieve state-of-the-art short term fractional frequency stability in CSOs in the past\cite{Hartnett2006apl,Locke2006rsi,DickWang1}.  Our results are the first observation of temperature turning points below the boiling point of liquid helium. Table \ref{turningpoints} lists the magic temperatures (turning points) measured for a range of WG modes. Operation of the WHIGMO at a millikelvin magic temperature rather than its current $\approx$8K would have the benefit of reduced thermal noise floor of the maser, with potential operation at the quantum limit. Operation at a millikelvin turning point, where the thermal coefficient $\alpha$ is vanishing, would minimise the effects from the considerable heating due to the large ($>$10 dBm) input power required to saturate the pump transition of the maser. This is similarly advantageous for CSOs \cite{Locke2006rsi} which typically circulate large amounts of power through the sapphire resonator, at least several milliwatt, while the cooling power at base temperature of a dilution refrigerator is only on the order of several hundred $\mu$W. Additionally, high power operation of quantum limited transducers could be attained at these temperatures.\\
\begin{table}[!phtb]
\caption{\label{turningpoints}Measured `magic temperatures' for a range of quasi-transverse magnetic WG modes. Several modes close to the Fe$^{3+}$ centre frequency exhibited strong distortion and could not be accurately tracked to determine the turning point. The $m=17$ mode required large power to excite and could not be measured below 80mK due to heating effects.}
\begin{tabular}{|c|c|}
	\hline
	\textbf{\textit{m}}	&	\textbf{Magic Temperature (mK)} \\
	\hline
	\hline
	13	&	89.75 \\
	\hline
	14	&	96.25 \\
	\hline
	15	&	Not trackable \\
	\hline
	16	&	Not trackable \\
	\hline
	17	&	Possible turnover below 80 mK \\
	\hline
	18	&	2749.35 \\
	\hline
	19	&	2280.30	\\
	\hline
	20	&	2089.75	\\
	\hline
\end{tabular}
\end{table}

In summary, we report the first characterisation of a single-crystal sapphire resonator at temperatures more than an order of magnitude lower than previously achieved, the first measurement of thermal bistability in a microwave sapphire resonator at these temperatures, and the first observation of millikelvin frequency-temperature turning points. We give a model for the thermal bistability threshold power, and show that it is closely dependent on the material properties of the sapphire resonator. We propose several reasons the effect has not been previously observed in this system. Typically the CSO/WHIGMO is operated very close to a frequency-temperature turning point where the temperature coefficient $\alpha$ is small, and the threshold power for bistability becomes significantly larger than normal operational power levels. In the experiments reported in this paper, the resonator was operated at low temperature, far from a turning point where the temperature coefficient was large, leading to a lower and readily observable threshold power. Frequency-temperature turning points as low as tens of millikelvin were observed for WG modes in the resonator, and we show that the thermal bistability effect can be suppressed by operating at these ``magic temperatures''. Additionally, mode $Q$-factors remained high and were comparable to their usual values at higher temperature, ruling out the existence of extra loss mechanisms in sapphire in the millikelvin regime. We conclude that single-crystal sapphire is an excellent candidate for both clock applications, and quantum metrology of macroscopic systems at temperatures approaching absolute zero.

\begin{acknowledgments}
This work was supported by the Australian Research Council and a University of Western Australia collaboration grant.
\end{acknowledgments}

\end{document}